\documentclass[journal]{IEEEtran}
\usepackage{mathrsfs}
\usepackage{cite}
\usepackage{amsmath}
\usepackage{amssymb}
\usepackage{stfloats}
\usepackage{graphicx}
\usepackage{setspace}
\usepackage{xcolor}
\usepackage{multirow}
\usepackage{slashbox}
\ifCLASSOPTIONcompsoc
\usepackage[tight,normalsize,sf,SF]{subfigure}
\else
\usepackage[tight,footnotesize]{subfigure}
\fi

\begin{document}
\title{Low-Complexity Recursive Convolutional Precoding for OFDM-based Large-Scale Antenna Systems}

\author{Yinsheng~Liu, Geoffrey Ye Li, Wei Han, and Zhangdui Zhong.
\thanks{This work is partly supported by general financial grant from the China postdoctorl science foundation (No. 2015M570924), start-up project of state key lab (No. RCS2015ZQ004), and fundamental research funds of central universities (No. 2015RC035 and No. 2016JBM003).}
\thanks{Yinsheng Liu and Zhangdui Zhong are with the School of Computer Science and Information Technology and State Key Laboratory of Rail Traffic Control and Safety, Beijing Jiaotong University, Beijing 100044, China, e-mail: \{ys.liu@bjtu.edu.cn, zhdzhong@bjtu.edu.cn\}}% <-this % stops a space
\thanks{Geoffrey Ye Li is with ITP lab, school of ECE, Georgia Institute of Technology, Atlanta 30313, Georgia, USA, e-mail: liye@ece.gatech.edu.}
\thanks{Wei Han is with Huawei Technologies, Co. Ltd., China, e-mail:wayne.hanwei@huawei.com.}
}

\maketitle
%\doublespacing

\begin{abstract}
%\boldmath
\emph{Large-scale antenna} (LSA) has gained a lot of attention recently since it can significantly improve the performance of wireless systems. Similar to \emph{multiple-input multiple-output} (MIMO) \emph{orthogonal frequency division multiplexing} (OFDM) or MIMO-OFDM, LSA can be also combined with OFDM to deal with frequency selectivity in wireless channels. However, such combination suffers from substantially increased complexity proportional to the number of antennas in LSA systems. For the conventional implementation of LSA-OFDM, the number of \emph{inverse fast Fourier transform}s (IFFTs) increases with the antenna number since each antenna requires an IFFT for OFDM modulation. Furthermore, \emph{zero-forcing} (ZF) precoding is required in LSA systems to support more users, and the required matrix inversion leads to a huge computational burden. In this paper, we propose a low-complexity recursive convolutional precoding to address the issues above. The traditional ZF precoding can be implemented through the recursive convolutional precoding in the time domain so that only one IFFT is required for each user and the matrix inversion can be also avoided. Simulation results show that the proposed approach can achieve the same performance as that of ZF but with much lower complexity.

\end{abstract}

\begin{IEEEkeywords}
Large-scale antenna, massive MIMO, precoding, OFDM.
\end{IEEEkeywords}

%\newpage
\section{Introduction}
By installing hundreds of antennas at the \emph{base station }(BS), \emph{large-scale antenna} (LSA) systems can significantly improve performance of cellular networks \cite{EGLarsson,FRusek}. Even if LSA can be regarded as an extension of the traditional \emph{multiple-input multiple-output} (MIMO) systems, which has been widely studied during the last couple of decades \cite{GJFoschini}, many special properties of LSA due to extremely large number of antennas make it a potential technique for future wireless systems and thus has gained lots of attention recently.\par

When the antenna number is sufficiently large, the performance in an LSA system becomes deterministic \cite{HQNgo}. From the power scaling law for LSA \cite{HQNgo}, the transmit power of each user is inversely proportional to the antenna number or the square root of the antenna number, depending on whether accurate channel state information is available or not. For downlink transmission with multiple users, precoding techniques are required at the BS to achieve the system capacity \cite[Ch. 10]{DTse}. When the antenna number is large enough and the channels corresponding to different antennas or users are independent, the channel vectors for different users are asymptotically orthogonal. If the user number is much smaller than the antenna number which is always true in LSA systems, the \emph{matched filter} (MF) will perform as well as the typical linear precoders, such as \emph{zero-forcing }(ZF) or \emph{minimum mean-square-error} (MMSE). Therefore, the complexity can be greatly reduced since no matrix inversion is required for precoding\cite{LLu,EGLarsson}.\par

Similar to the philosophy of MIMO-\emph{orthogonal frequency division multiplexing} (OFDM) \cite{LJCimini} or MIMO-OFDM \cite{YLi_99}, LSA can be also combined with OFDM to deal with frequency selectivity in wireless channels. Although straightforward, such combination suffers from substantially increased complexity.\par

First, the precoding is conducted in the frequency domain for traditional MIMO-OFDM  \cite{3GPP_series}. In this case, each antenna at the BS requires an \emph{inverse fast Fourier transform} (IFFT) for OFDM modulation and the number of IFFTs is equal to the antenna number. Therefore, the number of IFFTs will increase substantially as the rising of the antenna number in LSA systems, leading to a huge computational burden.\par

Second, \emph{zero-forcing} (ZF) precoding is required to support more users in LSA systems. As indicated in \cite{EGLarsson,FRusek}, the MF precoding can perform as well as the ZF precoding in LSA systems because the \emph{inter-user-interference} (IUI) can be suppressed asymptotically through the MF precoding if the antenna number is large enough and the channels at different antennas and different users are independent. In practical systems, however, the antenna number is always finite. Moreover, the channels at different antennas will be correlated when placing so many antennas in a small area. In this sense, there will be residual IUI for the MF precoding, and the ZF precoding is thus still required \cite{JHoydis}. As a result, the matrix inversion of the ZF precoding will substantially increase the complexity, especially when the user number is large.\par

To address the issues above, we propose a low-complexity recursive convolutional precoding for LSA-OFDM in this paper.\par

First, a convolutional precoding filter in the time domain is used to replace the traditional precoding in the frequency domain. In this way, only one IFFT is required for each user no matter how many antennas there are. Meanwhile, by exploiting the frequency-domain correlation of the traditional precoding coefficients, the length of the precoding filter can be much smaller than the FFT size. As a result, the complexity can be greatly reduced, especially when the antenna number is large. Even though the convolutional precoding has been studied in \cite{YWLiang} for traditional MIMO-OFDM systems, its advantage is not as significant as in LSA systems. In this paper, we highlight that such advantage becomes remarkable when the antenna number is large and thus it is more suitable to adopt the convolutional precoding rather than the traditional frequency-domain precoding for the transceiver design in LSA-OFDM systems.

Second, based on the order recursion of Taylor expansion, the convolutional precoding filter works recursively in this paper such that we can not only avoid direct matrix inverse of traditional ZF precoding but also provide a way to implement the traditional ZF precoding through the convolutional precoding filter with low complexity. Taylor expansion has already been used for \emph{Truncated polynomial expansion} (TPE) in \cite{AMuller,AKammoun,NShariati,GMASessler}. In \cite{AMuller}, it is used to approximate the matrix inverse in ZF precoding. The precoding can be conducted iteratively so that the matrix inverse can be avoided. A similar approach is adopted in \cite{AKammoun} where the TPE is based on Cayley-Hamilton theorem and Taylor expansion is used for optimization of polynomial coefficients. The order recursion of Taylor expansion has also been used in \cite{NShariati,GMASessler} for channel estimation and multiuser detection. Different from the existing works that are based on a matrix form Taylor expansion in the frequency domain, the recursive ZF precoding in this paper is implemented through the recursive filter in the time domain such that it can be naturally combined with the convolutional precoding. Moreover, the order recursion is converted to a time recursion in this paper so that the proposed approach can track the time-variation of channels. Based on the time recursion, the tracking property is further analyzed for large-scale regime, resulting in new theoretical insights for the behaviors of time recursion in LSA systems that are not revealed before.\par

The rest of this paper is organized as follows. The system model is introduced in Section II. The proposed approach is derived in Section III, and its performance is analyzed in Section III. Simulation results are presented Section V. Finally, conclusions are drawn in Section VI.

\section{System Model}
Consider downlink transmission in an LSA-OFDM system where a BS employs $M$ antennas to serve $P$ users, each with one antenna, simultaneously at the same frequency band. As in \cite{EGLarsson}, we assume $M\gg P$.\par
Denote $x_p[n,k]$ with $\mathrm{E}(|x_p[n,k]|^2)=E_s$ to be the transmit symbol for the $p$-th user at the $k$-th subcarrier of the $n$-th OFDM block. In an LSA-OFDM based on traditional OFDM implementation, the precoding is carried out in the frequency domain, and therefore the transmit signal at the $l$-th sample of the $n$-th OFDM block at the $m$-th antenna for the $p$-th user is
\begin{align}
s_{m,p}[n,l]=\frac{1}{\sqrt{K}}\sum_{k=0}^{K-1}u_{m,p}[n,k]x_p[n,k]e^{j\frac{2\pi kl}{K}},
\end{align}
where $K$ denotes the subcarrier number for the OFDM modulation and $u_{m,p}[n,k]$ denotes the precoding coefficient for the $k$-th subcarrier of the $n$-th OFDM block at the $m$-th antenna for the $p$-th user. A \emph{cyclic prefix }(CP) will be added in front of the transmit signal to deal with the delay spread of wireless channels.\par
After removing the CP and OFDM demodulation, the received signal at the $p$-th user can be expressed as
\begin{align}\label{2-2}
y_p[n,k]=\sum_{m=1}^Mh_{p,m}[n,k]\left(\sum_{p=1}^Pu_{m,p}[n,k]x_p[n,k]\right)+z_p[n,k],
\end{align}
where $z_p[n,k]$ is the additive white noise with $\mathrm{E}(|z_p[n,k]|^2)=N_0$, and $h_{p,m}[n,k]$ is the \emph{channel frequency response} (CFR) corresponding to the $k$-th subcarrier of the $n$-th block at the $m$-th antenna for the $p$-th user, which can be expressed as
\begin{align}
h_{p,m}[n,k]=\sum_{l=0}^{L-1}c_{p,m}[n,l]e^{-j\frac{2\pi lk}{K}},
\end{align}
where $c_{p,m}[n,l]$ is the \emph{channel impulse response} (CIR) and $L$ denotes the channel length which is usually much smaller than the FFT size. The CFR is assumed to be complex Gaussian distributed with zero mean and $\mathrm{E}\{h_{p,m}[n,k]h_{p_1,m_1}^*[n,k]\}=g_p\rho[m - m_1]\delta[p-p_1]$, where $g_p$ denotes the square of the large-scale fading coefficient for the $p$-th user, $\rho[\cdot]$ denotes the correlation function of the channels at different antennas for the same user, and $\delta[\cdot]$ denotes the Kronecker delta function. It means the CFRs have been assumed to be independent for different users while they depend on the correlation function, $\rho[\cdot]$, for different antennas. In particular, we have $\rho[\cdot]=\delta[\cdot]$ when the CFRs at different antennas are independent.\par

From (\ref{2-2}), the received signal vector corresponding to the $k$-th subcarrier of the $n$-th OFDM block for all users can be expressed as
\begin{align}\label{2-1}
\mathbf{y}[n,k]&\triangleq(y_1[n,k],\cdots,y_P[n,k])^{\mathrm{T}}\nonumber\\
&=\mathbf{H}[n,k]\mathbf{U}[n,k]\mathbf{x}[n,k]+\mathbf{z}[n,k],
\end{align}
where
\begin{align}
\mathbf{x}[n,k]&=(x_1[n,k],\cdots,x_P[n,k])^{\mathrm{T}},\nonumber\\
\mathbf{z}[n,k]&=(z_1[n,k],\cdots,z_P[n,k])^{\mathrm{T}},\nonumber\\
\mathbf{U}[n,k]&=\{u_{m,p}[n,k]\}_{m,p=1}^{M,P}=(\mathbf{u}_1[n,k],\cdots,\mathbf{u}_P[n,k]),\nonumber\\
\mathbf{H}[n,k]&=\{h_{p,m}[n,k]\}_{p,m=1}^{P,M}=\left(\mathbf{h}_1[n,k],\cdots,\mathbf{h}_P[n,k]\right)^{\mathrm{T}},\nonumber
\end{align}
with $\mathbf{u}_p[n,k]=(u_{1,p}[n,k],\cdots,u_{M,p}[n,k])^{\mathrm{T}}$ being the corresponding precoding vector of the $p$-th user and $\mathbf{h}_p[n,k]=\left(h_{p,1}[n,k],\cdots,h_{p,M}[n,k]\right)^{\mathrm{T}}$ being the CFR vector for the $p$-th user with correlation matrix $\mathrm{E}\{\mathbf{h}_p[n,k]\mathbf{h}_p^{\mathrm{H}}[n,k]\}\triangleq g_p\mathbf{R}$ where $\{\mathbf{R}\}_{(m,m_1)}=\rho[m-m_1]$.\par

\section{Low-Complexity Recursive Convolutional Precoding}
In this section, we will first present recursive updating of precoding matrices, then derive the low-complexity convolutional precoding, and discuss its complexity at the end of this section.
\subsection{Recursive Updating}
The ZF precoding is considered in this paper although the proposed approach can be also used for other precodings, such as the MMSE precoding. Assuming the downlink channels are known at the BS, the desired precoding matrix can be expressed as
\begin{align}\label{3-1}
\mathbf{U}_o[n,k]=\mathbf{H}^{\mathrm{H}}[n,k]\left(\mathbf{H}[n,k]\mathbf{H}^{\mathrm{H}}[n,k]\right)^{-1}.
\end{align}\par
Using Taylor expansion in Appendix A, the matrix inverse in (\ref{3-1}) can be substituted by an order-recursive relation as
\begin{align}\label{3-2}
&\mathbf{U}^{(Q+1)}[n,k]=\mathbf{U}^{(Q)}[n,k]+\nonumber\\
&\frac{\mu}{M}\mathbf{H}^{\mathrm{H}}[n,k]\mathbf{G}^{-1}(\mathbf{I}-\mathbf{H}[n,k]\mathbf{U}^{(Q)}[n,k]),
\end{align}
where $\mathbf{G}=\mathrm{diag}\{g_p\}_{p=1}^P$ and $\mathbf{U}^{(Q)}[n,k]$ denotes the corresponding precoding matrix with the $Q$-th order expansion and $\mu$ is a step size that affects the convergence, as we will discuss in Section IV. The order-recursive relation in (\ref{3-2}) can be also rewritten in a vector form as
\begin{align}\label{3-3}
&\mathbf{u}_p^{(Q+1)}[n,k]=\mathbf{u}_p^{(Q)}[n,k]+\nonumber\\
&\frac{\mu}{M}\sum_{i=1}^Pg_i^{-1}\mathbf{h}_i^*[n,k](\delta[i-p]-\mathbf{h}_i^{\mathrm{T}}[n,k]\mathbf{u}_p^{(Q)}[n,k]),
\end{align}
where $\mathbf{u}_p^{(Q)}[n,k]$ denotes the $p$-th column of $\mathbf{U}^{(Q)}[n,k]$.\par
In (\ref{3-3}), the order-recursive updating is driven by the expansion order, $Q$. Mathematically, the expansion order in (\ref{3-3}) can be viewed as a \emph{recursion counter}, which increases as the recursion proceeds. In this sense, the OFDM block index can be also used as that \emph{recursion counter}. In other words, (\ref{3-3}) can be also driven by the OFDM block index if replacing expansion order, $Q$, with OFDM block index, $n$, that is
\begin{align}\label{3-4}
&\mathbf{u}_p[n+1,k]=\mathbf{u}_p[n,k]+\nonumber\\
&\frac{\mu}{M}\sum_{i=1}^Pg_i^{-1}\mathbf{h}_i^*[n,k]({\delta[i-p]-\mathbf{h}_i^{\mathrm{T}}[n,k]\mathbf{u}_p[n,k]}).
\end{align}
As a result, the order recursion in (\ref{3-3}) is converted to the time recursion in (\ref{3-4}). Essentially, the order recursion in (\ref{3-3}) can be converted to the time recursion in (\ref{3-4}) is just because they have a similar expression except that one is driven by $Q$ and the other is driven by $n$. Using the time recursion in (\ref{3-4}), the actual calculation can be conducted in the time domain even though the principle for avoiding the matrix inverse is based on the order recursion in (\ref{3-3}). In this way, we can not only reduce the complexity since there is not need to repeat the order recursions from the zeroth order for each OFDM block, but also track the time-varying channels as long as the channel changes slowly. Strictly speaking, the above conversion is only valid when the channel is time invariant. In this case, (\ref{3-3}) and (\ref{3-4}) have exactly the same expression except for different \emph{recursion counter}s. In practice, the time recursion in (\ref{3-4}) can still work as long as the channel is slowly time-varying. Our analysis in Section IV shows that the time recursion can track the time variation of the channels when Doppler frequency is small but the performance will degrade as the rising of Doppler frequency.

\begin{figure}
  \center
  \includegraphics[width=3in]{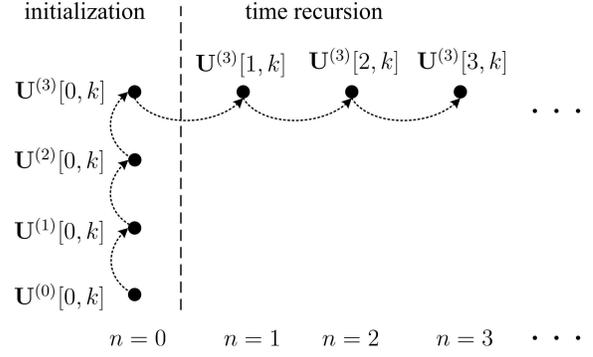}\\
  \caption{The order recursion is used for initialization and the time recursion is used for updating the coefficients of the sequent OFDM blocks.}\label{hybrid}
\end{figure}

Actually, the order recursion and the time recursion can be used in a hybrid manner as in Fig.~\ref{hybrid}. The order recursion is used for initialization and the time recursion is used for tracking. Once the expansion order for initialization is large enough to achieve satisfied performance, the time recursion will be on to update the precoding coefficients in the subsequent OFDM blocks. In this way, we can save the complexity since only one recursion is needed to update the coefficients during the tracking stage.

\begin{figure*}
  \centering
  \includegraphics[angle=90,width=3.5in]{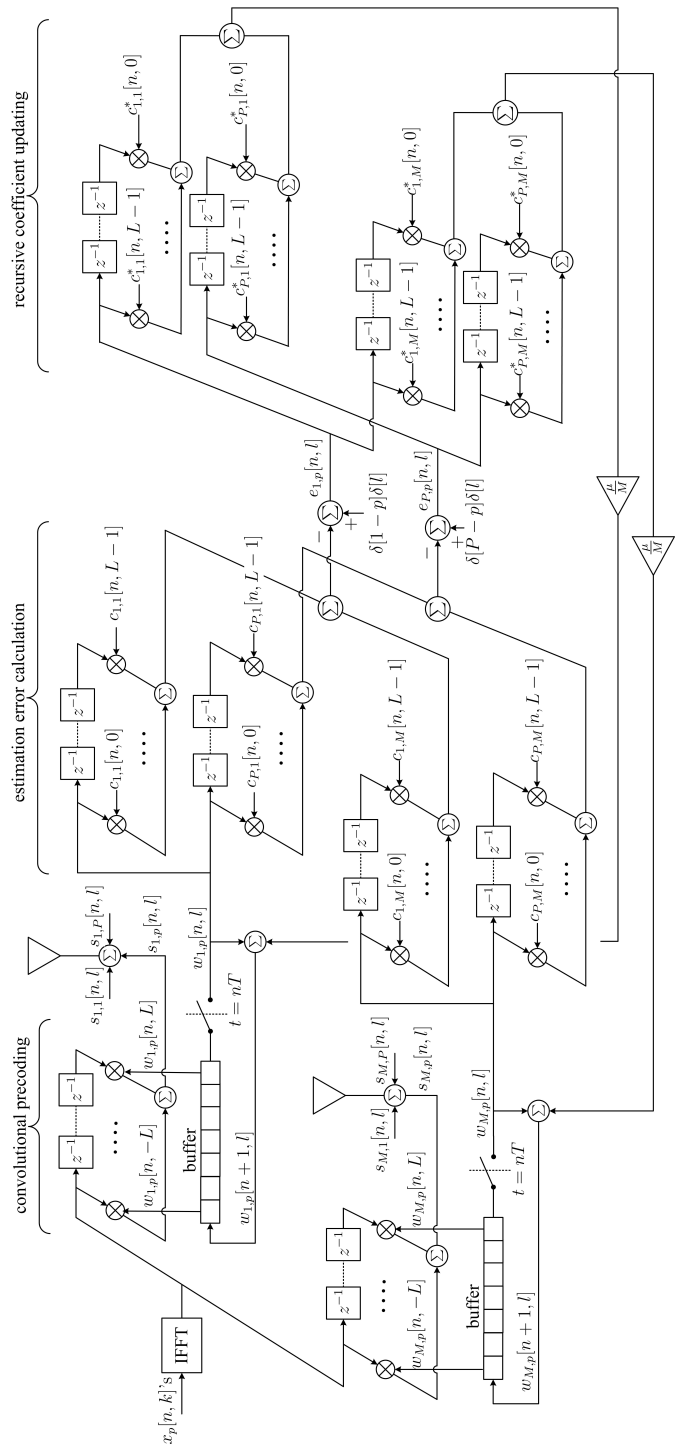}
  \caption{Recursive convolutional precoding with convolutional precoding, recursive coefficient updating, and estimation error calculation.}\label{structure}
\end{figure*}

\subsection{Convolutional Precoding}
Although the matrix inverse is avoided through (\ref{3-4}), the precoding is still conducted in the frequency domain. In this subsection, we will convert it into the time-domain convolutional precoding by exploiting the frequency-domain correlation of the precoding matrices. Denote $\mathbf{u}_{m,p}[n]=(u_{m,p}[n,0],\cdots,u_{m,p}[n,K-1])^{\mathrm{T}}$, which contains the precoding coefficients from all subcarriers of the $n$-th OFDM block at the $m$-th antenna for the $p$-th user. Then, (\ref{3-4}) can be rewritten as
\begin{align}\label{3-5}
&\mathbf{u}_{m,p}[n+1]=\mathbf{u}_{m,p}[n]+\nonumber\\
&\frac{\mu}{M}\sum_{i=1}^{P}g_i^{-1}\left({\delta[i-p]\mathbf{I}}-\mathbf{D}_{i,p}[n]\right)\mathbf{h}^*_{i,m}[n],
\end{align}
where $\mathbf{h}_{p,m}[n]=(h_{p,m}[n,0],\cdots,h_{p,m}[n,K-1])^{\mathrm{T}}$ is the corresponding CFR vector from the $m$-th antenna to the $p$-th user, and $\mathbf{D}_{i,p}[n]$ is a $K\times K$ diagonal matrix with the $(k,k)$-th element given by
\begin{align}
\{\mathbf{D}_{i,p}[n]\}_{(k,k)}=\sum_{m=1}^M{h}_{i,m}[n,k]u_{m,p}[n,k].
\end{align}
Denote $w_{m,p}[n,l]$ to be the coefficient for the $l$-th tap of the precoding filter at the $m$-th antenna for the $p$-th user corresponding to the $n$-th OFDM block. Then, we have
\begin{align}
\mathbf{w}_{m,p}[n]&\triangleq(w_{m,p}[n,0],\cdots,w_{m,p}[n,K-1])^{\mathrm{T}}\nonumber\\
&=\frac{1}{K}\mathbf{F}^{\mathrm{H}}\mathbf{u}_{m,p}[n],
\end{align}
where $\mathbf{w}_{m,p}[n]$ is the corresponding precoding vector, and $\mathbf{F}$ is the \emph{discrete Fourier transform} (DFT) matrix with the $(m,n)$-th element given by
\begin{align}\label{B2}
\{\mathbf{F}\}_{(m,n)}=e^{-j\frac{2\pi m n}{K}},~~~~m,n\in[0,K-1].
\end{align}
From Appendix B, by taking the inverse DFT of (\ref{3-5}), we can obtain the coefficients for the time-domain convolutional precoding filter as
\begin{align}\label{3-6}
w_{m,p}[n+1,l]=w_{m,p}[n,l]+\frac{\mu}{M}\sum_{i=1}^{P}g_i^{-1}c_{i,m}^*[n,-l] * e_{i,p}[n,l],
\end{align}
where $e_{i,p}[n,l]$ is the estimation error given by
\begin{align}
e_{i,p}[n,l]=\delta[i-p]\delta[l]-\sum_{m=1}^Mc_{i,m}[n,l] *  w_{m,p}[n,l].
\end{align}\par

The resulted recursive convolutional precoding is shown in Fig.~\ref{structure}, where large-scale fading is omitted by setting $g_p=1$. The precoding is carried out in the time domain via the precoding filter. In this case, only one IFFT is required for each user no matter how many antennas there are at the BS. Therefore, the number of IFFTs is equal to the number of users, which is much smaller than the antenna number in LSA systems. By exploiting the correlation of frequency-domain precoding coefficients, the coefficients of the precoding filter is sparse and thus can be truncated. For the single user case, the precoding filter is exactly the conjugate of the CIR and thus $0\leq l\leq L-1$. In the case of multiple users, we use one more tap, as a rule of thumb, for the positive taps and another $L$ taps to include the significant coefficients on the negative taps. As a result, $w_{m,p}[n,l]$ can be truncated within the range $-L\leq l\leq L$ (modulo $K$). Following the order recursion based initialization, the coefficients of the precoding filter can be updated recursively.\par

Note that the transmit signal after the IFFT should be circularly extended before sending to the precoding filter so that the signal can be circularly convolved with the precoding filter because the production in the frequency domain corresponds to the circular convolution in the time domain \cite{AVOppenheim}.

\subsection{Complexity}

\begin{table*}
\caption{Comparison of Complexities for convolutional precoding and traditional ZF precoding.}\centering
\begin{tabular}{|c||c|c|c|}
  \hline
  \multirow{2}{*}{\backslashbox{Complexity}{Approaches}} & \multirow{2}{*}{Proposed} & \multirow{2}{*}{Traditional ZF} & \multirow{2}{*}{TPE} \\
  & &  & \\
  \hline
  \multirow{2}{*}{IFFT} & \multirow{2}{*}{${\frac{1}{2}}PK\log_2K$} & \multirow{2}{*}{${\frac{1}{2}}MK\log_2K$} & \multirow{2}{*}{${\frac{1}{2}}MK\log_2K$}\\
  & & & \\
  \hline
  \multirow{2}{*}{Precoding operation} & \multirow{2}{*}{$PM(2L+1)$} & \multirow{2}{*}{$PMK$} & \multirow{2}{*}{$PMK(2Q-1)$}\\
  & & &\\
  \hline
  \multirow{2}{*}{Coefficient calculation}  & \multirow{2}{*}{$2P^2ML$}  & \multirow{2}{*}{$\frac{1}{B}(2P^2MK+\mathcal{O}(P^3)K)$} & \multirow{2}{*}{$-$} \\
  & & &\\
  \hline
\end{tabular}\label{tab}
\end{table*}

\begin{figure}
  \centering
  \includegraphics[width=3.5in]{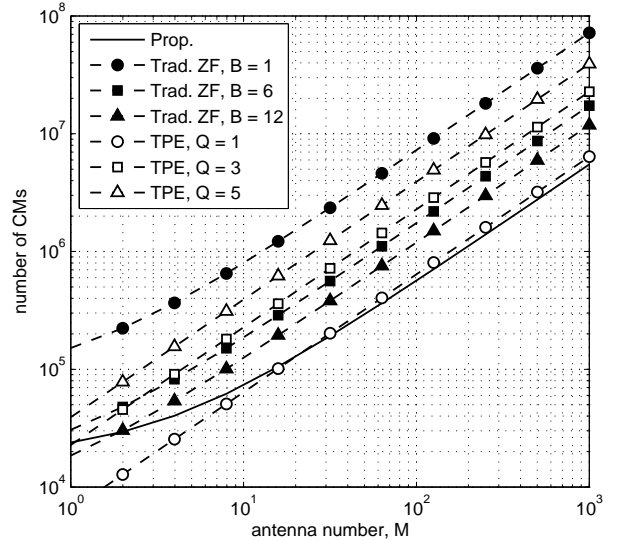}\\
  \caption{An example for the complexity comparison for $P=8$ user case.}\label{complexity}
\end{figure}

In Tab.~\ref{tab}, the complexity is evaluated in terms of the number of \emph{complex multiplications} (CMs) required by the IFFT, the actual precoding operation, and the coefficient calculation for the precoding \cite[Ch. 2]{TSauer}. As comparisons, the complexities of the traditional ZF precoding and the TPE precoding in \cite{AMuller} are also included in the table. For the traditional ZF precoding, $B$ consecutive subcarriers ($B=12$ in \emph{long-term evolution }(LTE)) can share the same precoding coefficients by exploiting the frequency-domain correlation of the precoding coefficients. For the TPE precoding, it requires $Q-1$ iterations for each OFDM block because the iterations are repeated from the zeroth order for each OFDM block.\par

We have the following observations from the table. First, the number of IFFTs is equal to the antenna number for the traditional ZF precoding and the TPE precoding, and thus the number of IFFTs for the proposed approach is greatly reduced since the user number is much smaller than the antenna number in LSA systems. Second, the precoding filter length for the convolutional precoding is much smaller than the FFT size, while the precoding operation has to be conducted on each subcarrier individually for the traditional ZF precoding and the TPE precoding. Third, the number of CMs can be reduced for the proposed approach because the coefficient calculation is conducted recursively, while the traditional ZF precoding can also reduce the number of CMs since $B$ consecutive subcarriers can have the same precoding coefficients.\par

As an example, Fig.~\ref{complexity} presents the CMs required by the proposed approach, the traditional ZF precoding, and the TPE precoding for the typical $5$ MHz bandwidth in LTE where the size of FFT is $K=512$ \cite{3GPP}. For a typical \emph{extended typical urban} (ETU) channel whose maximum delay $\tau_{\mathrm{max}} = 5\mu\text{s}$, a channel length $L=38$ is enough to contain most of the channel power. As expected, the complexity of the convolutional precoding is substantially reduced compared with existing approaches when the antenna number is large. When antenna number is small, however, the complexity reduction is not so significant as that for the case of large antenna number. The traditional ZF or TPE may even require fewer CMs than the proposed approach with larger $B$ or smaller $Q$, at the cost of performance degradation, as will be shown in Section V. In fact, the advantage of the convolutional precoding can be hardly observed in traditional systems since the antenna number there is small, and it only becomes remarkable when the antenna number is very large. Therefore, it is more suitable to adopt the convolutional precoding rather than the traditional frequency-domain precoding for the transceiver design in LSA-OFDM systems. Note that the convolutional precoding will cause some delay of the signal transmission. However, the complexity reduction is favorable if the delay due to the convolution is tolerable.

\section{Performance Analysis}
In this section, we will first analyze the convergence performances of initialization and tracking, respectively, and then discuss the impacts of imperfect channels. Since the time-domain convolutional precoding is equivalent to the frequency-domain precoding, the performance analysis is conducted in the frequency domain for simplicity.
\subsection{Initialization}

We focus on the OFDM block with $n=0$ where the order-recursion is used for initialization. Define $\Delta{\mathbf{U}}^{(Q)}[0,k]\mathbf{G}^{\frac{1}{2}}\triangleq \mathbf{U}_o[0,k]\mathbf{G}^{\frac{1}{2}}-\mathbf{U}^{(Q)}[0,k]\mathbf{G}^{\frac{1}{2}}$ to be the normalized precoding matrix error for initialization, where the large-scale fading effect has been taken into account. Then it is shown in Appendix C that
\begin{align}\label{3-6_1}
\|\Delta{\mathbf{U}}^{(Q)}[0,k]\mathbf{G}^{\frac{1}{2}}\|_{\mathrm{F}}^2=\frac{1}{M}\sum_{p=1}^P\lambda_p^{-1}(1-\mu\lambda_p)^{2(Q+1)},
\end{align}
where $\lambda_p$ is the $p$-th eigenvalue of $\frac{1}{M}\mathbf{H}^{\mathrm{H}}[0,k]\mathbf{G}^{-1}\mathbf{H}[0,k]$ or $\frac{1}{M}\mathbf{G}^{-\frac{1}{2}}\mathbf{H}[0,k]\mathbf{H}[0,k]^{\mathrm{H}}\mathbf{G}^{-\frac{1}{2}}$.\par

Denote $\lambda_{\mathrm{max}}$ and $\lambda_{\mathrm{min}}$ to be the maximum and the minimum eigenvalues of $\frac{1}{M}\mathbf{H}^{\mathrm{H}}[0,k]\mathbf{G}^{-1}\mathbf{H}[0,k])$, respectively. From (\ref{3-6_1}), the convergence can be achieved as long as $0<\mu<2/\lambda_{\mathrm{max}}$, and the optimal step size for the fastest convergence will be $\mu_0=2/(\lambda_{\mathrm{max}}+\lambda_{\mathrm{min}})$ \cite{SHaykin}. Depending on whether the channels at different antennas are independent or not, we have the following discussions:
\begin{itemize}
\item
If the CFRs corresponding to different antennas are independent, we have $\lambda_p\approx 1$ for $p=1,2,\cdots,P$ \cite[Cha. 1]{ZBai}. In this case, fast convergence can be achieved by setting $\mu_o= 1$, and the convergence can be almost achieved within only one recursion as we can see from the simulation results in the next section.

\item
If the CFRs corresponding to different antennas are correlated, the maximum and the minimum eigenvalues will rely on $\mathbf{G}^{-\frac{1}{2}}\mathbf{H}[0,k]\mathbf{H}[0,k]^{\mathrm{H}}\mathbf{G}^{-\frac{1}{2}}$. Inspired by $\mathrm{E}\{\mathbf{G}^{-\frac{1}{2}}\mathbf{H}[0,k]\mathbf{H}[0,k]^{\mathrm{H}}\mathbf{G}^{-\frac{1}{2}}\}=P\mathbf{R}$, we let $\lambda_{\mathrm{max}}=\lambda_{\mathrm{max}}{(\mathbf{R})}$ and $\lambda_{\mathrm{min}}=\lambda_{\mathrm{min}}{(\mathbf{R})}$ for simplicity, and thus $\mu_0=2/[\lambda_{\mathrm{max}}(\mathbf{R}) + \lambda_{\mathrm{min}}(\mathbf{R})]$ in this case. Obviously, such step size can cover the case where the channels at different antennas are independent because $\mathbf{R}=\mathbf{I}$ in that situation. Simulation results in Section V shows such step size can work well for the proposed approach.
\end{itemize}
\subsection{Tracking}
When the channel is static, the performance of tracking will be the same with that in (\ref{3-6_1}) except that the expansion order, $Q$, is replaced by the block index, $n$. On the other hand, if the channel is time-varying, the variation of the desired precoding matrix is given, from (\ref{3-1}), by
\begin{align}\label{3-8_1}
\mathbf{\Phi}[n,k]&\triangleq\mathbf{U}_o[n+1,k]-\mathbf{U}_o[n,k]\nonumber\\
&=\mathbf{H}^{\mathrm{H}}[n+1,k]\left(\mathbf{H}[n+1,k]\mathbf{H}^{\mathrm{H}}[n+1,k]\right)^{-1}-\nonumber\\
&~~~~\mathbf{H}^{\mathrm{H}}[n,k]\left(\mathbf{H}[n,k]\mathbf{H}^{\mathrm{H}}[n,k]\right)^{-1}.
\end{align}\par
Exact analysis based on (\ref{3-8_1}) is difficult. To gain analytical insights, we assume the channels corresponding to different antennas and different users are independent. In that case, $\mathbf{\Phi}[n,k]$ can be approximated by
\begin{align}
\mathbf{\Phi}[n,k]\approx\frac{1}{M}\left(\mathbf{H}^{\mathrm{H}}[n+1,k]-\mathbf{H}^{\mathrm{H}}[n,k]\right)\mathbf{G}^{-1}.
\end{align}
Furthermore, we assume that the expansion order for initialization is sufficiently large so that $\mathbf{U}[0,k]=\mathbf{U}_o[0,k]$.\par

Define $\Delta{\mathbf{U}}[n,k]\mathbf{G}^{\frac{1}{2}}\triangleq\mathbf{U}_o[n,k]\mathbf{G}^{\frac{1}{2}}-\mathbf{U}[n,k]\mathbf{G}^{\frac{1}{2}}$ to be the normalized precoding matrix error for tracking, where the large-scale fading effect has been taken into account. When the Doppler frequency, $f_d$, is small, then it is shown in Appendix D that the \emph{mean-square-error } (MSE) can be expressed by
\begin{align}\label{3-8_3}
\mathrm{MSE}_n(M,P)&\triangleq\mathrm{E}\{\|\Delta{\mathbf{U}}[n,k]\mathbf{G}^{\frac{1}{2}}\|_{\mathrm{F}}^2\}\nonumber\\
&=\frac{2\pi^2 f_d^2 T^2 M}{P}\left[1-\left(1-\frac{P}{M}\right)^n\right]^2,
\end{align}
where $T$ denotes the OFDM symbol duration. From (\ref{3-8_3}), we have the following observations:
\begin{itemize}
\item The MSE of tracking depends only on the ratio of user number and antenna number. As the antenna number is much larger than the user number in an LSA system, we have
    \begin{align}
    \mathrm{MSE}_n(M,P)\approx 2\pi^2f_d^2T^2n^2\frac{P}{M}.
    \end{align}
\item The MSE of tracking increases as the rising of OFDM block index. It means that the performance will be degraded as the time recursion proceeds which can be also confirmed by our simulation results.

\item The MSE of tracking increases as the rising of Doppler frequency, that is, the performance will be degraded as the rising of Doppler frequency, which also coincides with our intuition.
\end{itemize}

\subsection{Impact of Imperfect Channel}
In the above, we have assumed that the accurate downlink channel is known at the BS. In practical systems, the downlink channel at the BS can be obtained by estimating the uplink channel due to the reciprocity in time-division duplexing systems \cite{survey}. In any case, only imperfect channel is known at the BS.\par

To analyze the impacts of channel estimation error, denote the imperfect channel to be
\begin{align}\label{3-9}
\widehat{\mathbf{H}}[n,k]=\mathbf{H}[n,k]+\widetilde{\mathbf{H}}[n,k],
\end{align}
where $\widetilde{\mathbf{H}}[n,k]=\{h_{p,m}[n,k]\}_{p,m=1}^{P,M}$ denotes the channel estimation error with $\mathrm{E}\{\widetilde{h}_{p,m}[n,k]\widetilde{h}_{p_1,m_1}^*[n,k]\}=g_p\sigma_h^2\delta[p-p_1]\delta[m-m_1]$ with $\sigma_h^2$ being the variance of the error when $g_p=1$. Assuming the CFRs and the channel errors are independent, we can obtain, when the antenna number is large enough, that,
\begin{align}\label{3-9_0}
\frac{1}{M}\widehat{\mathbf{H}}[n,k]\widehat{\mathbf{H}}^{\mathrm{H}}[n,k]\approx \frac{1}{M}\mathbf{H}[n,k]\mathbf{H}^{\mathrm{H}}[n,k]+\sigma_h^2\mathbf{G}.
\end{align}
From (\ref{3-9_0}), we have $\widehat{\lambda}_{p}=\lambda_p+\sigma_h^2$ where $\widehat{\lambda}_p$ denotes the $p$-th eigenvalue of $\frac{1}{M}\widehat{\mathbf{H}}^{\mathrm{H}}[n,k]\mathbf{G}^{-1}\widehat{\mathbf{H}}[n,k]$ or $\frac{1}{M}\mathbf{G}^{-\frac{1}{2}}\widehat{\mathbf{H}}[n,k]\widehat{\mathbf{H}}[n,k]^{\mathrm{H}}\mathbf{G}^{-\frac{1}{2}}$. For simplicity, we will only focus on the initialization in the subsequential of this subsection, although our results are also available for the tracking stage.\par

In the presence of the channel estimation error, the order recursion for initialization can be rewritten by
\begin{align}\label{3-9_1}
&\widehat{\mathbf{U}}^{(Q+1)}[0,k]=\widehat{\mathbf{U}}^{(Q)}[0,k]+\nonumber\\
&\frac{\mu}{M}\widehat{\mathbf{H}}^{\mathrm{H}}[0,k]\mathbf{G}^{-1}(\mathbf{I}-\widehat{\mathbf{H}}[0,k]\widehat{\mathbf{U}}^{(Q)}[0,k]),
\end{align}
where $\widehat{\mathbf{U}}^{(Q)}[0,k]$ denotes the precoding coefficients with imperfect channel. Correspondingly, the normalized precoding matrix error is $\Delta{\widehat{\mathbf{U}}}^{(Q)}[0,k]\mathbf{G}^{\frac{1}{2}}\triangleq\widehat{\mathbf{U}}_o[0,k]\mathbf{G}^{\frac{1}{2}}-\widehat{\mathbf{U}}^{(Q)}[0,k]\mathbf{G}^{\frac{1}{2}}$ where $\widehat{\mathbf{U}}_o[0,k]=\widehat{\mathbf{H}}^{\mathrm{H}}[0,k](\widehat{\mathbf{H}}[0,k]\widehat{\mathbf{H}}^{\mathrm{H}}[0,k])^{-1}$ indicates the desired precoding matrix with imperfect channel. Following the same analysis in Section IV.A, the convergence of (\ref{3-9_1}) can be achieved by choosing $\mu_o=1+\sigma_h^2$ when the channels at different antennas are independent or $\mu_o=2/[\lambda_{\mathrm{max}}(\mathbf{R})+\lambda_{\mathrm{min}}(\mathbf{R}) + 2\sigma_h^2]$ when they are correlated. We have $\|\Delta{\widehat{\mathbf{U}}}^{(\infty)}[0,k]\mathbf{G}^{\frac{1}{2}}\|_{\mathrm{F}}^2=0$ and thus ${\widehat{\mathbf{U}}}^{(\infty)}[0,k]=\widehat{\mathbf{U}}_o[0,k]$ when $Q\rightarrow\infty$.\par
In addition to changing the step size, the channel estimation error will also cause the performance degradation when the convergence has been achieved. Denote $\Delta \mathbf{U}_o[0,k]\mathbf{G}^{\frac{1}{2}}\triangleq \mathbf{U}_o[0,k]\mathbf{G}^{\frac{1}{2}}-\widehat{\mathbf{U}}_o[0,k]\mathbf{G}^{\frac{1}{2}}$ to be the error for the desired precoding matrix due to the channel estimation error. To gain analytical insights, we assume the channels corresponding to different antennas and different users are independent. In that case,
\begin{align}\label{3-9_2}
\Delta \mathbf{U}_o[0,k]\mathbf{G}^{\frac{1}{2}}&\approx\frac{1}{M}\mathbf{H}^{\mathrm{H}}[0,k]\mathbf{G}^{-\frac{1}{2}}-\frac{1}{M(1+\sigma_h^2)}\widehat{\mathbf{H}}^{\mathrm{H}}[0,k]\mathbf{G}^{-\frac{1}{2}}\nonumber\\
&=\frac{1}{M(1+\sigma_h^2)}(\sigma_h^2\mathbf{H}^{\mathrm{H}}[0,k]-\widetilde{\mathbf{H}}^{\mathrm{H}}[0,k])\mathbf{G}^{-\frac{1}{2}}.
\end{align}
With the assumption that the CFRs and the channel errors are independent, the MSE can be expressed by
\begin{align}\label{3-9_3}
\mathrm{E}\{\|\Delta \mathbf{U}_o[0,k]\mathbf{G}^{\frac{1}{2}}\|_{\mathrm{F}}^2\}=\frac{P\sigma_h^2}{M(1+\sigma_h^2)}.
\end{align}
From (\ref{3-9_3}), we have the following observations:
\begin{itemize}
\item If assuming $\sigma_h^2$ is very small, we have
\begin{align}
\mathrm{E}\{\|\Delta \mathbf{U}_o[0,k]\mathbf{G}^{\frac{1}{2}}\|_{\mathrm{F}}^2\}\approx \frac{P\sigma_h^2}{M},
\end{align}
which is approximately proportional to the variance of the channel estimation error.
\item By increasing the antenna number or reducing the user number, the impact of the channel estimation error can be mitigated. In the extreme case where $M\rightarrow\infty$, we have $\mathrm{E}\{\|\Delta \mathbf{U}_o[0,k]\mathbf{G}^{\frac{1}{2}}\|_{\mathrm{F}}^2\}=0$, which means the impact of the channel estimation error vanishes when the antenna number is very large.
\end{itemize}

\section{Simulation Results}

\begin{figure}
  \centering
  \subfigure{
  \includegraphics[width=3.5in]{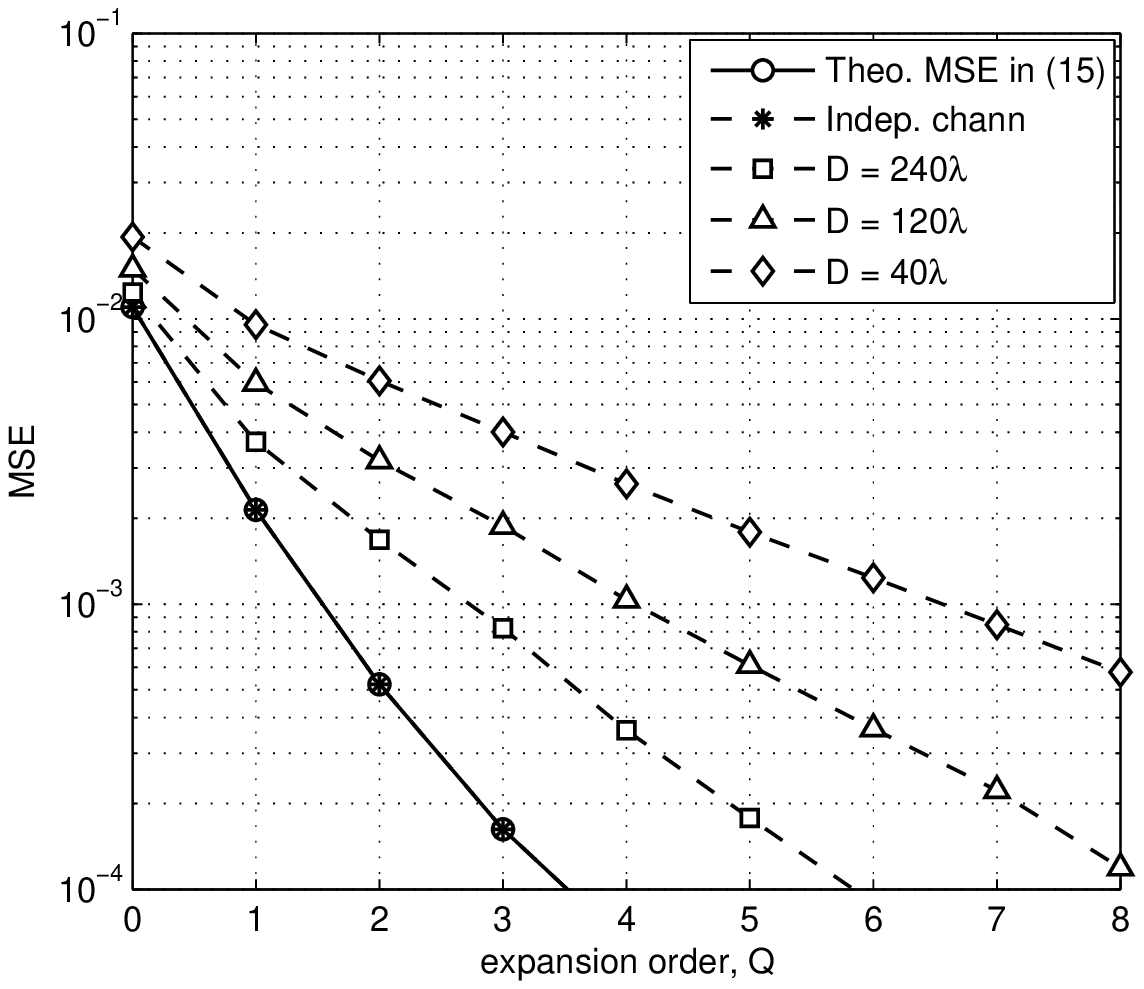}
  }\\
  (a)\\
  \subfigure{
  \includegraphics[width=3.5in]{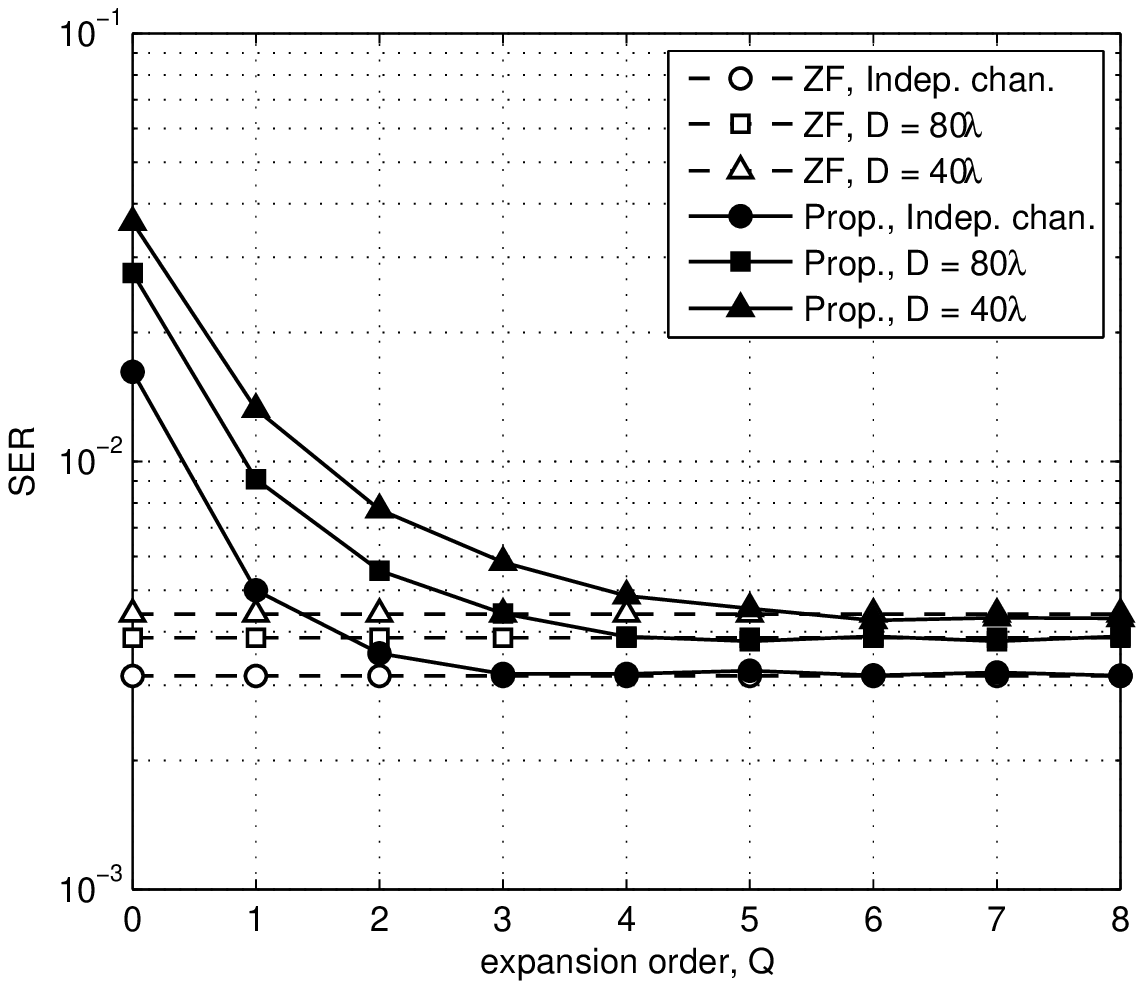}
  }\\
  (b)\\
  \caption{Performances of initialization (a) MSE (b) SER.}\label{initialization}
\end{figure}

\begin{figure}
  \centering
  \subfigure{
  \includegraphics[width=3.5in]{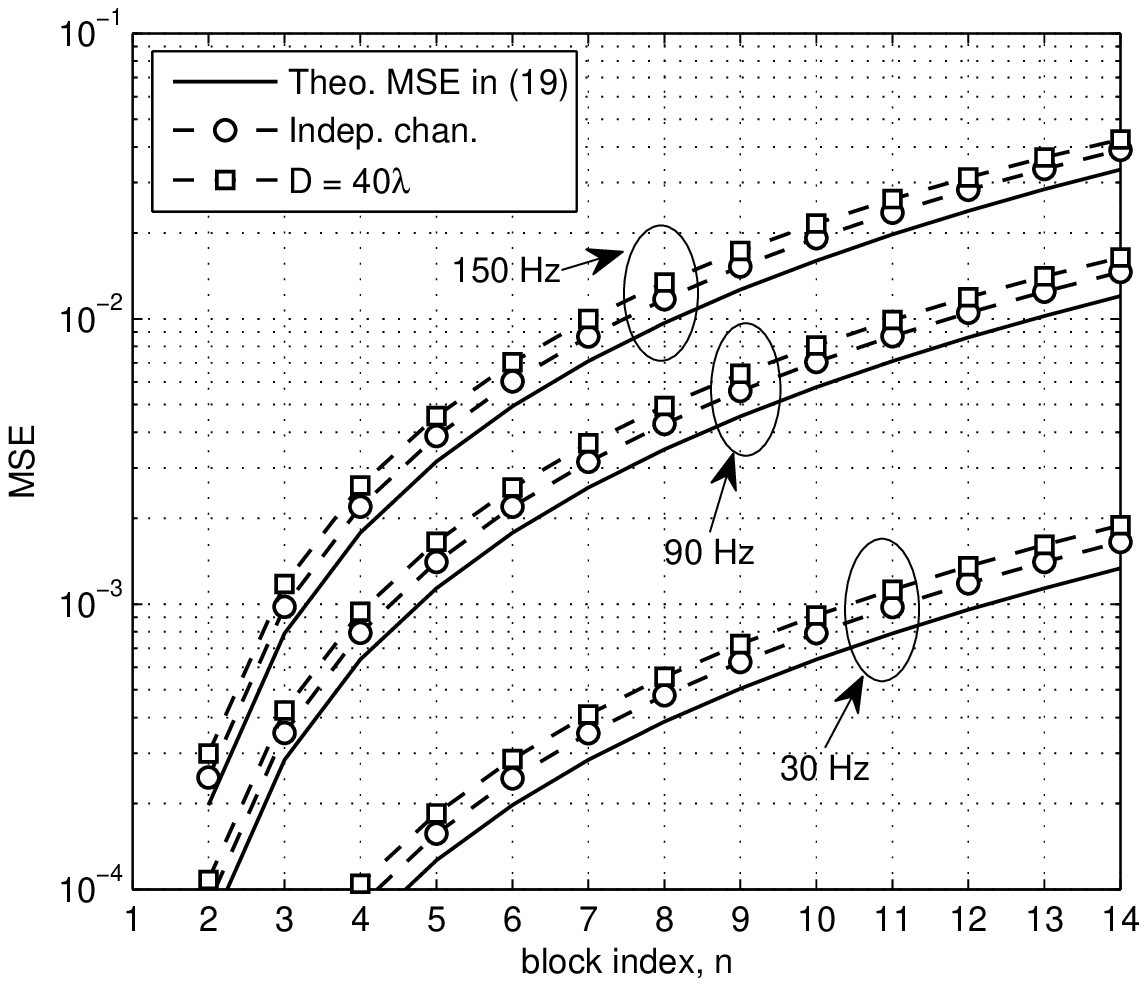}
  }\\
  (a)\\
  \subfigure{
  \includegraphics[width=3.5in]{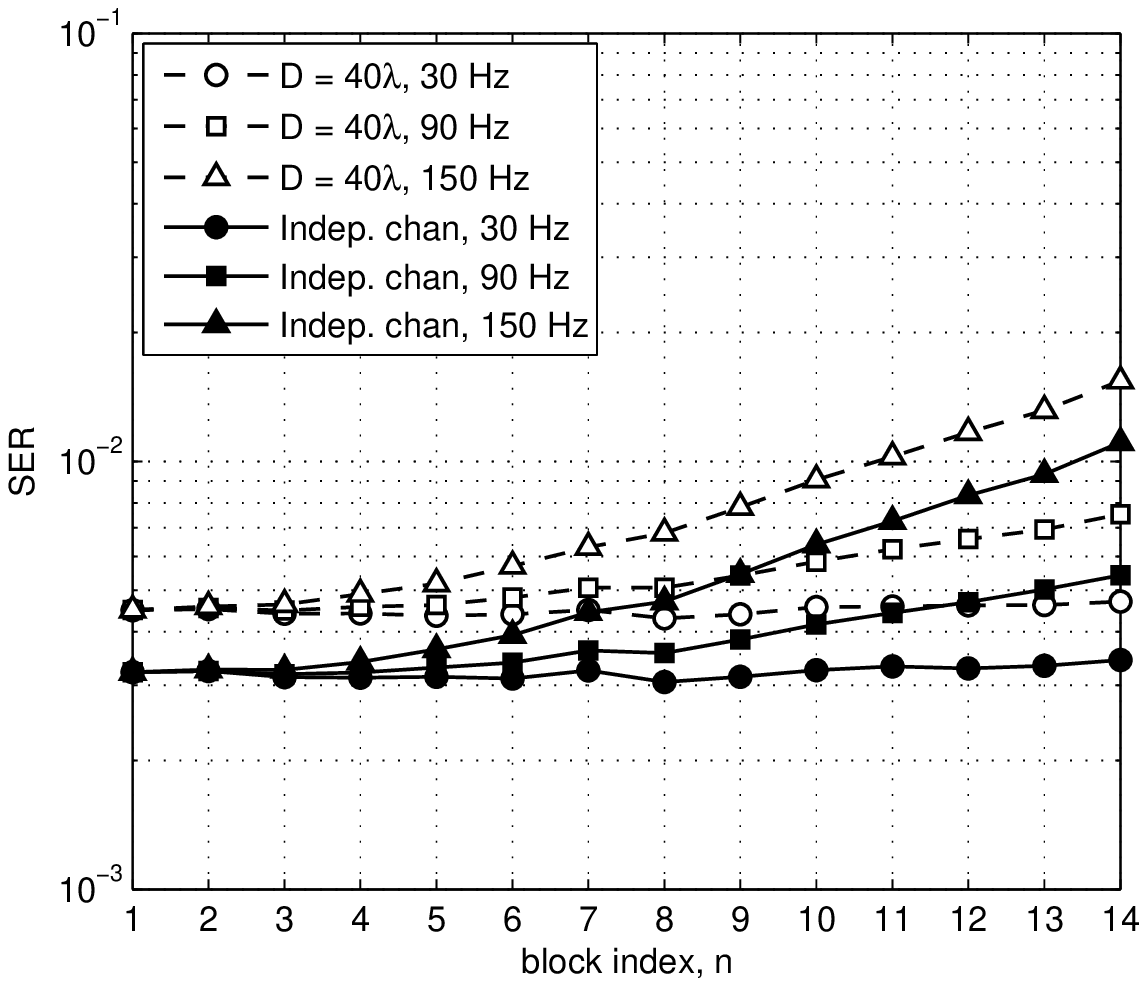}
  }\\
  (b)\\
  \caption{Performances of tracking (a) MSE (b) SER.}\label{tracking}
\end{figure}

\begin{figure}
  \centering
  \subfigure{
  \includegraphics[width=3.5in]{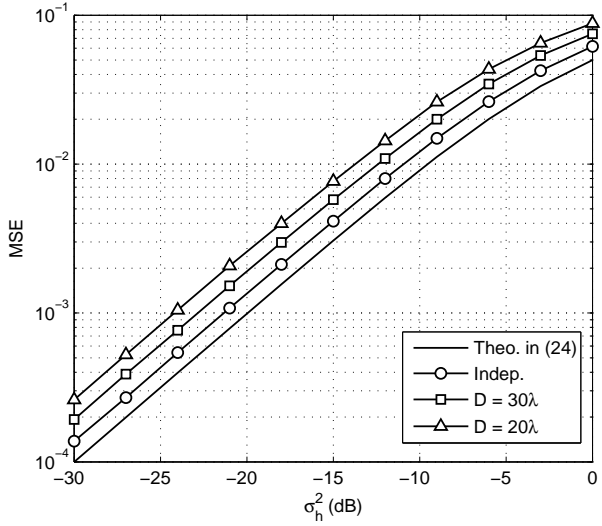}
  }\\
  (a)\\
  \subfigure{
  \includegraphics[width=3.5in]{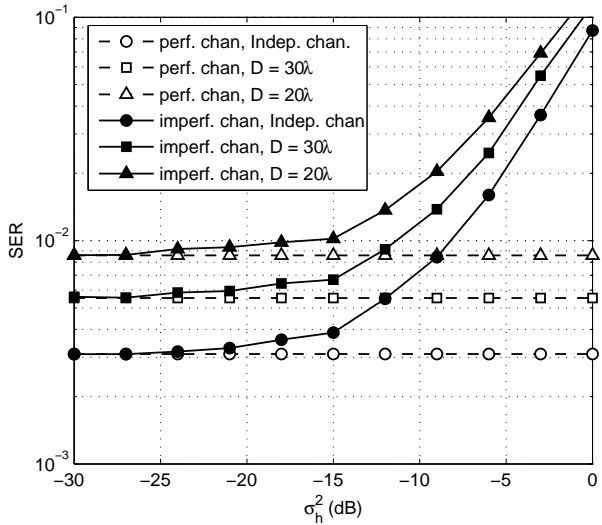}
  }\\
  (b)\\
  \caption{Impacts of imperfect channel information for (a) MSE (b) SER.}\label{chan_error}
\end{figure}

\begin{figure}
  \centering
  \includegraphics[width=3.5in]{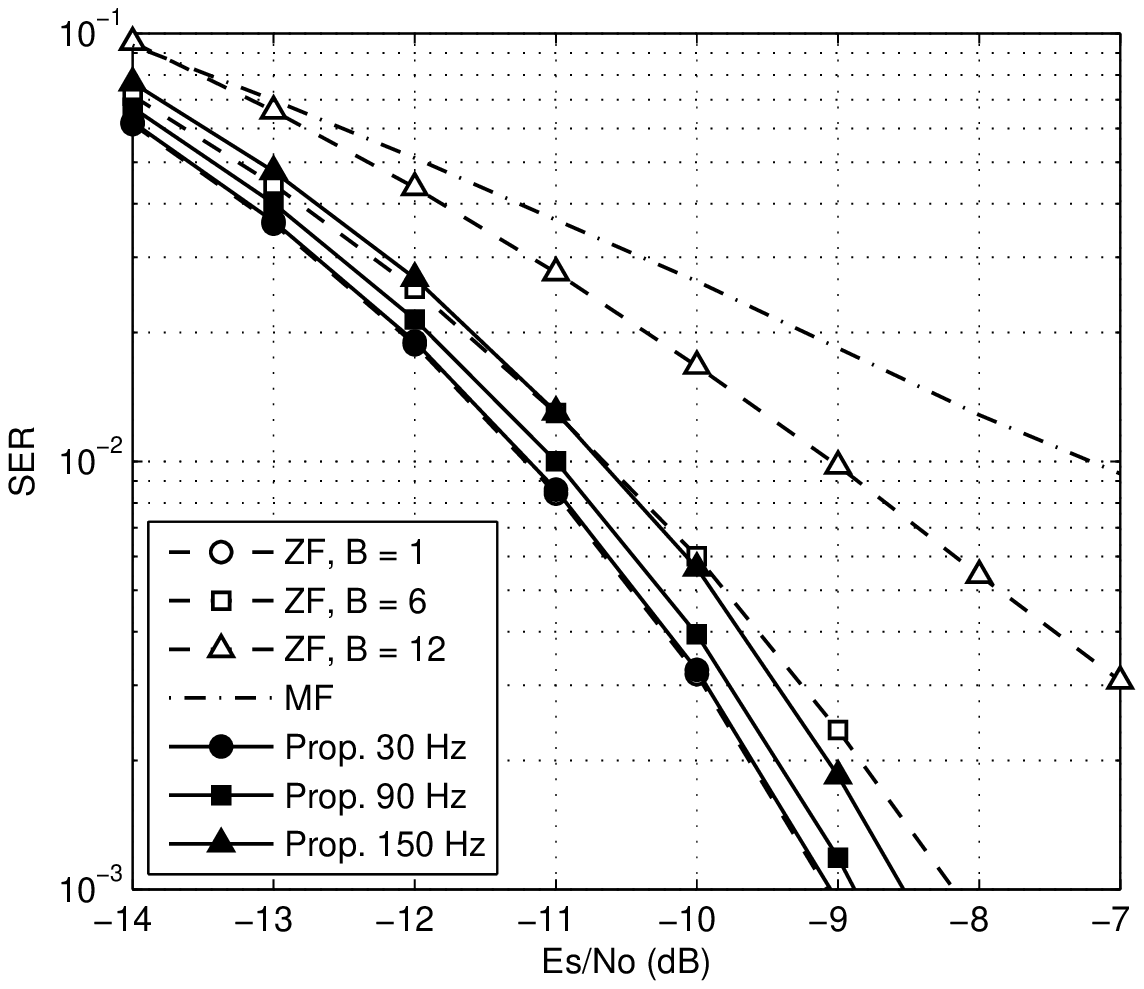}\\
  \caption{SER versus $Es/N_0$.}\label{SERvsSNR}
\end{figure}

In this section, we evaluate the proposed approach using computer simulation. We consider a BS equipped with $M=100$ antennas and $P=10$ users in the system. A \emph{quadrature-phase-shift-keying} (QPSK) modulated OFDM signal is used, where the subcarrier spacing is $15$ KHz corresponding to an OFDM symbol duration about $66.7\mu\text{s}$. For a typical $5$ MHz channel, the size of FFT is $512$ with $300$ subcarriers used for data transmission and the others used as guard band as in LTE \cite{3GPP}. Each frame consists of $14$ OFDM symbols. A normalized ETU channel model is used, which has $9$ taps and the maximum delay $\tau_{\mathrm{max}} = 5\mu\text{s}$. The channels at different antennas can be independent or correlated. For the latter, a \emph{uniform-linear-array} (ULA) is used where the antennas are placed along a straight line \cite{YLiu2}. In this case, the correlation of channels at $m$-th antenna and $m_1$-th antenna is $\rho[m-m_1]=J_0[2\pi (m - m_1) D/(M-1)]$, where $D$ is the array size normalized by the wavelength. Apparently, the channels at different antennas will be more correlated for smaller $D$. Without loss of generality, we assume $g_p=1$ for all users.\par

Fig.~\ref{initialization} shows the MSE and \emph{symbol-error-ratio} (SER) for the initialization of the proposed approach. From Fig.~\ref{initialization} (a), the MSE reduces as the order recursion proceeds. However, the MSEs for the correlated channels cannot reduce as fast as that for the independent channels. It means that more order recursions are required to achieve a satisfied performance for the initialization when the channels at different antennas are correlated. This coincides with the observation in Fig.~\ref{initialization} (b). From Fig.~\ref{initialization} (b), the SER can be improved as the order recursion proceeds. When the channels at different antennas are independent, the proposed approach can achieve the same SER with the ZF precodings within only two recursions. However, more recursions are required when the channels at different antennas are correlated.\par

Fig.~\ref{tracking} shows the MSE and SER for the tracking of the proposed approach with different Doppler frequencies. We assume the expansion order for initialization is large enough such that $\mathbf{U}[0,k]=\mathbf{U}_o[0,k]$. From Fig.~\ref{tracking} (a), the channel correlation causes smaller impact to the tracking MSE than it does to the initialization MSE. From Fig.~\ref{tracking} (b), the time-varying channels can be efficiently tracked when the Doppler frequency is small and therefore the SERs over different OFDM blocks will be almost the same. On the other hand, it becomes difficult to track the channel time variation as the increasing of the Doppler frequency, and thus the SERs for the OFDM blocks at the end of the frame will get worse. This problem can be easily addressed by re-initialization when the precoding coefficients are getting far from the desired ones.\par

Fig.~\ref{chan_error} shows the impacts of the channel estimation error. From (\ref{chan_error}) (a), the MSE is approximately proportional to the variance of the channel estimation error when the latter is small, which coincides with our analysis in Section IV. Fig.~\ref{chan_error} (b) shows that the channel estimation error has little affects on the SER when $\sigma_h^2<-15$ dB. Otherwise, the SER performances will be seriously degraded as the increasing of the channel estimation error.\par

Fig.~\ref{SERvsSNR} shows the SER versus $Es/N_0$ with different Doppler frequencies. For the proposed approach, we also assume the expansion order is large enough for initialization such that $\mathbf{U}[0,k]=\mathbf{U}_o[0,k]$. As the increasing of the Doppler frequencies, the SER performances degrade because the channels cannot be efficiently tracked when the Doppler frequency is large. As comparisons, the MF precoding and the traditional ZF precodings with $B=1,6,12$ are also included. Since the ZF and MF precodings are conducted for each OFDM block individually, the SER performances will be the same for different Doppler frequencies. When the Doppler frequency is small, the proposed approach can achieve the same SER as the traditional ZF precoding with $B=1$. As the increasing of $B$, the performance of ZF precoding will degrade although the complexity can be reduced. Meanwhile, the proposed approach can significantly outperform the MF precoding since the latter cannot completely remove the IUI.

\section{Conclusions}
In this paper, low-complexity convolutional precoding has been proposed for the precoder design in an LSA-OFDM system. The traditional frequency-domain precoding has been converted into a time-domain convolutional precoding so that the number of IFFTs is substantially reduced. On the other hand, based on the order recursion of Taylor expansion, the convolutional precoding filter works recursively in this paper such that we can not only avoid direct matrix inverse of traditional ZF precoding but also provide a way to implement the traditional ZF precoding through the convolutional precoding filter with low complexity. Our results have shown that it is more suitable to adopt the convolutional precoding rather than the traditional frequency-domain precoding for the transceiver design in LSA-OFDM systems.

\appendices
\renewcommand{\theequation}{A.\arabic{equation}}
\setcounter{equation}{0}
\section{Derivation of (\ref{3-2})}
When the antenna number is sufficiently large and the CFRs corresponding to different users and different antennas are independent, the CFR vectors for different users are asymptotically orthogonal and therefore we have
\begin{align}
\frac{1}{M}\mathbf{H}[n,k]\mathbf{H}^{\mathrm{H}}[n,k]=\mathbf{G}.
\end{align}
In practical systems, however, the antenna number is always finite, and the channels at different antennas can be correlated when placing so many antennas in a small area. In such case,
\begin{align}\label{A1}
\frac{1}{M}\mathbf{H}[n,k]\mathbf{H}^{\mathrm{H}}[n,k]=\mathbf{G}-\mathbf{\Delta}[n,k],
\end{align}
where $\mathbf{\Delta}[k]$ can be viewed as a perturbation matrix. When scaled by a factor $\mu$, we have
\begin{align}\label{A2}
\frac{\mu}{M}\mathbf{H}[n,k]\mathbf{H}^{\mathrm{H}}[n,k]\mathbf{G}^{-1}=\mathbf{I}-\mathbf{\Lambda}[n,k],
\end{align}
where $\mathbf{\Lambda}[n,k]=(1-\mu)\mathbf{I}+\mu\mathbf{\Delta}[n,k]\mathbf{G}^{-1}$. Using the Taylor expansion, the inverse of (\ref{A2}) can be expressed by
\begin{align}\label{A3}
\mathbf{P}^{(Q)}[n,k]\triangleq\left(\mathbf{I}-\mathbf{\Lambda}[n,k]\right)^{-1}=\sum_{q=0}^Q\mathbf{\Lambda}^q[n,k].
\end{align}
Substituting (\ref{A3}) into (\ref{3-1}), we can obtain
\begin{align}\label{A4}
\mathbf{U}^{(Q)}[n,k]=\frac{\mu}{M}\mathbf{H}^{\mathrm{H}}[n,k]\mathbf{G}^{-1}\mathbf{P}^{(Q)}[n,k],
\end{align}
where $\mathbf{U}^{(Q)}[n,k]$ denotes the precoding matrix with the $Q$-th order Taylor expansion. Exploiting the relation between consecutive expansion orders, we have
\begin{align}\label{A5}
\mathbf{P}^{(Q+1)}[n,k]=\mathbf{I}+\mathbf{\Lambda}[n,k]\mathbf{P}^{(Q)}[n,k].
\end{align}
Substituting (\ref{A5}) into (\ref{A4}),
\begin{align}
&~~~~\mathbf{U}^{(Q+1)}[n,k]\nonumber\\
&=\frac{\mu}{M}\mathbf{H}^{\mathrm{H}}[n,k]\mathbf{G}^{-1}+\frac{\mu}{M}\mathbf{H}^{\mathrm{H}}[n,k]\mathbf{G}^{-1}\mathbf{\Lambda}[n,k]\mathbf{P}^{(Q)}[n,k]\nonumber\\
&=\frac{\mu}{M}\mathbf{H}^{\mathrm{H}}[n,k]\mathbf{G}^{-1}+\frac{\mu}{M}\mathbf{H}^{\mathrm{H}}[n,k]\mathbf{G}^{-1}\cdot\nonumber\\
&~~~~\left(\mathbf{I}-\frac{\mu}{M}\mathbf{H}[n,k]\mathbf{H}^{\mathrm{H}}[n,k]\mathbf{G}^{-1}\right)\mathbf{P}^{(Q)}[n,k]\nonumber\\
&=\frac{\mu}{M}\mathbf{H}^{\mathrm{H}}[n,k]\mathbf{G}^{-1}+\frac{\mu}{M}\mathbf{H}^{\mathrm{H}}[n,k]\mathbf{G}^{-1}\mathbf{P}^{(Q)}[n,k]-\nonumber\\
&~~~~\frac{\mu^2}{M^2}\mathbf{H}^{\mathrm{H}}[n,k]\mathbf{G}^{-1}\mathbf{H}[n,k]\mathbf{H}^{\mathrm{H}}[n,k]\mathbf{G}^{-1}\mathbf{P}^{(Q)}[n,k]\nonumber\\
&=\mathbf{U}^{(Q)}[n,k]+\frac{\mu}{M}\mathbf{H}^{\mathrm{H}}[n,k]\mathbf{G}^{-1}\left(\mathbf{I}-\mathbf{H}[n,k]\mathbf{U}^{(Q)}[n,k]\right).
\end{align}

\renewcommand{\theequation}{B.\arabic{equation}}
\setcounter{equation}{0}
\section{Derivation of (\ref{3-6})}
Taking the inverse DFT on both sides of (\ref{3-5}), we have
\begin{align}\label{B3}
&~~~~\mathbf{w}_{m,p}[n+1]\nonumber\\
&=\mathbf{w}_{m,p}[n]+\frac{\mu}{M}\sum_{i=1}^Pg_i^{-1}\frac{\mathbf{F}^{\mathrm{H}}}{\sqrt{K}}\cdot\nonumber\\
&~~~~\left({\delta[i-p]\mathbf{I}}-\mathbf{D}_{i,p}[n]\right)\frac{\mathbf{F}}{\sqrt{K}}\frac{1}{K}\mathbf{F}^{\mathrm{H}}\mathbf{h}_{i,m}^*[n]\nonumber\\
&=\mathbf{w}_{m,p}[n]+\frac{\mu}{M}\sum_{i=1}^Pg_i^{-1}\frac{\mathbf{F}^{\mathrm{H}}}{\sqrt{K}}\cdot\nonumber\\
&~~~~\left({\delta[i-p]\mathbf{I}}-\mathbf{D}_{i,p}[n]\right)\frac{\mathbf{F}}{\sqrt{K}}
\left(\begin{array}{c}
c_{i,m}^*[n,0] \\
\vdots \\
c_{i,m}^*[n,-(K-1)]
\end{array}
\right).
\end{align}
To proceed, we can derive that
\begin{align}\label{B4}
&~~~~\frac{\mathbf{F}^{\mathrm{H}}}{\sqrt{K}}\mathbf{D}_{i,p}[n]\frac{\mathbf{F}}{\sqrt{K}}\nonumber\\
&=\sum_{m_0=1}^M\frac{\mathbf{F}^{\mathrm{H}}}{\sqrt{K}}\mathrm{diag}\{{h}_{i,m_0}[n,k]u_{m_0,p}[n,k]\}\frac{\mathbf{F}}{\sqrt{K}}\nonumber\\
&=\sum_{m_0=1}^M\mathrm{circ}\{c_{i,m_0}[n,l]\circledast w_{m_0,p}[n,l]\},
\end{align}
where $\circledast$ denotes the circular convolution and $\mathrm{circ}\{a_0,a_1,\cdots,a_{K-1}\}$ denotes a circular matrix constructed using $a_0,a_1,\cdots,a_{K-1}$. Substituting (\ref{B4}) into (\ref{B3}), we have
\begin{align}\label{B5}
&~~~~\mathbf{w}_{m,p}[n+1]\nonumber\\
&=\mathbf{w}_{m,p}[n]+\frac{\mu}{M}\sum_{i=1}^Pg_i^{-1}\left[\delta[i-p]\left(\begin{array}{c}
c_{i,m}^*[n,0] \\
\vdots \\
c_{i,m}^*[n,-(K-1)]
\end{array}
\right)-\right.\nonumber\\
&~~\left.\sum_{m_0=1}^M\mathrm{circ}\{c_{i,m_0}[n,l]\circledast w_{m_0,p}[n,l]\}
\left(\begin{array}{c}
c_{i,m}^*[n,0] \\
\vdots \\
c_{i,m}^*[n,-(K-1)]
\end{array}
\right)\right],
\end{align}
which can be rewritten in a scalar form as
%\begin{align}\label{B6}
%w_{m,p}[n+1,l]&=w_{m,p}[n,l]+\frac{\mu}{M}\sum_{i=1}^Pg_i^{-1}\Bigg(\delta[i-p]c_{i,m}^*[n,-l]-\nonumber\\
%&~~\left.\sum_{m_0=1}^Mc_{i,m_0}[n,l] \circledast w_{m_0,p}[n,l] \circledast c_{i,m}^*[n,-l]\right),
%\end{align}
%or
\begin{align}\label{B7}
w_{m,p}[n+1,l]&=w_{m,p}[n,l]+\frac{\mu}{M}\sum_{i=1}^Pg_i^{-1}c_{i,m}^*[n,-l]\circledast\nonumber\\
&~~\left(\delta[i-p]\delta[l]-\sum_{m_0=1}^Mc_{i,m_0}[n,l] \circledast w_{m_0,p}[n,l]\right).
\end{align}\par
In general, the channel length, $L$, is much smaller than the FFT size. In other words, the power of CIR, $c_{i,m}[n,l]$, may concentrate only on the taps at the beginning and the others are small enough and thus can be omitted. This is also the case for the precoding coefficients, $w_{m,p}[n,l]$, due to the correlation of frequency-domain precoding matrices. As a result, the circular convolution in (\ref{B7}) can be replaced by the linear convolution, leading exactly to (\ref{3-6}).

\renewcommand{\theequation}{C.\arabic{equation}}
\setcounter{equation}{0}
\section{Derivation of (\ref{3-6_1})}
By subtracting $\mathbf{U}_o[0,k]$ on both sides of (\ref{3-2}) and then multiplying $\mathbf{G}^{\frac{1}{2}}$, we obtain
\begin{align}\label{C1}
\Delta{\mathbf{U}}^{(Q+1)}[0,k]\mathbf{G}^{\frac{1}{2}}=&\left(\mathbf{I}-\frac{\mu}{M}\mathbf{H}^{\mathrm{H}}[0,k]\mathbf{G}^{-1}\mathbf{H}[0,k]\right)\nonumber\\
&\Delta{\mathbf{U}}^{(Q)}[0,k]\mathbf{G}^{\frac{1}{2}},
\end{align}
Denote
\begin{align}\label{C2}
\frac{1}{\sqrt{M}}\mathbf{G}^{-\frac{1}{2}}\mathbf{H}[0,k]=\mathbf{W}\mathbf{\Sigma}\mathbf{V}^{\mathrm{H}}=\mathbf{W}\mathbf{\Sigma}_0\mathbf{V}_0^{\mathrm{H}},
\end{align}
where $\mathbf{W}$ is a $P\times P$ unitary matrix, $\mathbf{\Sigma}=\left(\mathbf{\Sigma}_0,\mathbf{0}\right)$ with $\mathbf{\Sigma}_0=\mathrm{diag}\{{\lambda_p}^{\frac{1}{2}}\}_{p=1}^P$ being a $P\times P$ diagonal matrix, and $\mathbf{V}=\left(\mathbf{V}_0,\mathbf{V}_1\right)$ is an $M\times M$ unitary matrix where $\mathbf{V}_0$ includes the first $P$ columns and $\mathbf{V}_1$ includes the last $M-P$ columns. Then, we have
\begin{align}\label{C3}
\frac{1}{M}\mathbf{H}^{\mathrm{H}}[0,k]\mathbf{G}^{-1}\mathbf{H}[0,k]=\mathbf{V}\mathbf{\Sigma}^{\mathrm{H}}\mathbf{\Sigma}\mathbf{V}^{\mathrm{H}}=\mathbf{V}_0\mathbf{\Sigma}_0^{\mathrm{H}}\mathbf{\Sigma}_0\mathbf{V}_0^{\mathrm{H}}.
\end{align}
Substituting (\ref{C3}) into (\ref{C1}), we can obtain
\begin{align}\label{C4}
\Delta{\mathbf{U}}^{(Q+1)}[0,k]\mathbf{G}^{\frac{1}{2}}=\mathbf{V}\left(\begin{array}{cc}
                                                      \mathbf{I}-\mu\mathbf{\Sigma}_0^{\mathrm{H}}\mathbf{\Sigma}_0 &  \\
                                                       & \mathbf{I}
                                                    \end{array}
\right)\mathbf{V}^{\mathrm{H}}\Delta{\mathbf{U}}^{(Q)}[0,k]\mathbf{G}^{\frac{1}{2}}.
\end{align}
Using the recursive relation in (\ref{C4}), we can derive that
\begin{align}\label{C5}
&~~~~\mathbf{V}^{\mathrm{H}}\Delta{\mathbf{U}}^{(Q)}[0,k]\mathbf{G}^{\frac{1}{2}}\nonumber\\
&=\left(\begin{array}{cc}
                                                      \left(\mathbf{I}-\mu\mathbf{\Sigma}_0^{\mathrm{H}}\mathbf{\Sigma}_0\right)^{Q} &  \\
                                                       & \mathbf{I}
                                                    \end{array}
\right)\left(\begin{array}{c}
               \mathbf{V}_0^{\mathrm{H}} \\
               \mathbf{V}_1^{\mathrm{H}}
             \end{array}
\right)\Delta{\mathbf{U}}^{(0)}[0,k]\mathbf{G}^{\frac{1}{2}}.
\end{align}\par

%\begin{align}\label{C5}
%&~~~~\mathbf{V}^{\mathrm{H}}\Delta{\mathbf{U}}^{(Q)}[0,k]\mathbf{G}^{\frac{1}{2}}\nonumber\\
%&=\left(\begin{array}{cc}
%                                                      \left(\mathbf{I}-\mu\mathbf{\Sigma}_0^{\mathrm{H}}\mathbf{\Sigma}_0\right)^{Q} &  \\
%                                                       & \mathbf{I}
%                                                    \end{array}
%\right)\mathbf{V}^{\mathrm{H}}\Delta{\mathbf{U}}^{(0)}[0,k]\mathbf{G}^{\frac{1}{2}}\nonumber\\
%&=\left(\begin{array}{cc}
%                                                      \left(\mathbf{I}-\mu\mathbf{\Sigma}_0^{\mathrm{H}}\mathbf{\Sigma}_0\right)^{Q} &  \\
%                                                       & \mathbf{I}
%                                                    \end{array}
%\right)\left(\begin{array}{c}
%               \mathbf{V}_0^{\mathrm{H}} \\
%               \mathbf{V}_1^{\mathrm{H}}
%             \end{array}
%\right)\Delta{\mathbf{U}}^{(0)}[0,k]\mathbf{G}^{\frac{1}{2}}.
%\end{align}\par
Recall that
\begin{align}\label{C6}
\mathbf{U}_0[0,k]&=\mathbf{H}^{\mathrm{H}}[0,k]\left(\mathbf{H}[0,k]\mathbf{H}^{\mathrm{H}}[0,k]\right)^{-1}\nonumber\\
&=\frac{1}{\sqrt{M}}\mathbf{V}_0\mathbf{\Sigma}_0^{-1}\mathbf{W}^{\mathrm{H}}\mathbf{G}^{-\frac{1}{2}},\\
\mathbf{U}^{(0)}[0,k]&=\frac{\mu}{M}\mathbf{H}^{\mathrm{H}}[0,k]\mathbf{G}^{-1}=\frac{\mu}{\sqrt{M}}\mathbf{V}_0\mathbf{\Sigma}_0^{\mathrm{H}}\mathbf{W}^{\mathrm{H}}\mathbf{G}^{-\frac{1}{2}}.
\end{align}
Therefore,
\begin{align}\label{C8}
\Delta{\mathbf{U}}^{(0)}[0,k]&=\mathbf{U}_0[0,k]-\mathbf{U}^{(0)}[0,k]\nonumber\\
&=\frac{1}{\sqrt{M}}\mathbf{V}_0\mathbf{\Sigma}_0^{-1}(\mathbf{I}-\mu\mathbf{\Sigma}_0^{\mathrm{H}}\mathbf{\Sigma}_0)\mathbf{W}^{\mathrm{H}}\mathbf{G}^{-\frac{1}{2}}.
\end{align}
As a result, we have $\mathbf{V}_1^{\mathrm{H}}\Delta{\mathbf{U}}^{(0)}[0,k]=\mathbf{0}$ since $\mathbf{V}_1^{\mathrm{H}}\mathbf{V}_0=\mathbf{0}$. Using this relation, (\ref{C5}) can be simplified as
\begin{align}\label{C9}
\mathbf{V}_0^{\mathrm{H}}\Delta{\mathbf{U}}^{(Q)}[0,k]\mathbf{G}^{\frac{1}{2}}&=\left(\mathbf{I}-\mu\mathbf{\Sigma}_0^{\mathrm{H}}\mathbf{\Sigma}_0\right)^{Q}
               \mathbf{V}_0^{\mathrm{H}}\Delta{\mathbf{U}}^{(0)}[0,k]\mathbf{G}^{\frac{1}{2}}\nonumber\\
               &=\frac{1}{\sqrt{M}}\mathbf{\Sigma}_0^{-1}\left(\mathbf{I}-\mu\mathbf{\Sigma}_0^{\mathrm{H}}\mathbf{\Sigma}_0\right)^{Q+1}\mathbf{W}^{\mathrm{H}}.
\end{align}
Therefore, $\|\Delta{\mathbf{U}}^{(Q)}[0,k]\mathbf{G}^{\frac{1}{2}}\|_{\mathrm{F}}^2$ can be expressed by
\begin{align}\label{C10}
\|\Delta{\mathbf{U}}^{(Q)}[0,k]\mathbf{G}^{\frac{1}{2}}\|_{\mathrm{F}}^2&=\|\mathbf{V}^{\mathrm{H}}\Delta{\mathbf{U}}^{(Q)}[0,k]\mathbf{G}^{\frac{1}{2}}\|_{\mathrm{F}}^2\nonumber\\
&=\|\mathbf{V}_0^{\mathrm{H}}\Delta{\mathbf{U}}^{(Q)}[0,k]\mathbf{G}^{\frac{1}{2}}\|_{\mathrm{F}}^2\nonumber\\
&=\left\|\frac{1}{\sqrt{M}}\mathbf{\Sigma}_0^{-1}\left(\mathbf{I}-\mu\mathbf{\Sigma}_0^{\mathrm{H}}\mathbf{\Sigma}_0\right)^{Q+1}\mathbf{W}^{\mathrm{H}}\right\|_{\mathrm{F}}^2\nonumber\\
&=\frac{1}{M}\sum_{p=1}^P\lambda_p^{-1}(1-\mu\lambda_p)^{2(Q+1)},
\end{align}
where the fact that $\lambda_p$ is a real number since $\mathbf{H}^{\mathrm{H}}[0,k]\mathbf{G}^{-1}\mathbf{H}[0,k]$ is a Hermite matrix has been used.

\renewcommand{\theequation}{D.\arabic{equation}}
\setcounter{equation}{0}
\section{Derivation of (\ref{3-8_3})}
To analyze the tracking performance, rewrite (\ref{3-4}) in a matrix form as
\begin{align}\label{D1}
\mathbf{U}[n+1,k]=\mathbf{U}[n,k]+\frac{1}{M}\mathbf{H}^{\mathrm{H}}[n,k]\mathbf{G}^{-1}(\mathbf{I}-\mathbf{H}[n,k]\mathbf{U}[n,k]),
\end{align}
where $\mu_o=1$ has been used. By subtracting the $\mathbf{U}_o[n+1,k]$ on both sides of (\ref{D1}) and multiplying $\mathbf{G}^{\frac{1}{2}}$ , we have
\begin{align}\label{D2}
\Delta{\mathbf{U}}[n+1,k]\mathbf{G}^{\frac{1}{2}}=\left(\mathbf{I}-\frac{1}{M}\mathbf{H}^{\mathrm{H}}[n,k]\mathbf{G}^{-1}\mathbf{H}[n,k]\right)\cdot\nonumber\\
\Delta{\mathbf{U}}[n,k]\mathbf{G}^{\frac{1}{2}}+\mathbf{\Phi}[n,k]\mathbf{G}^{\frac{1}{2}},
\end{align}
which is a random differential equation whose system matrix is $\mathbf{I}-\frac{1}{M}\mathbf{H}^{\mathrm{H}}[n,k]\mathbf{G}^{-1}\mathbf{H}[n,k]$ \cite[Ch. 5]{SHaykin}. Due to low-pass filter effect of \emph{least-mean-square} (LMS) filter, we can adopt the direct-averaging method so that the instantaneous system matrix can be replaced by an average system matrix \cite{HJKushner},
\begin{align}\label{DC3}
\mathrm{E}\left\{\mathbf{I}-\frac{1}{M}\mathbf{H}^{\mathrm{H}}[n,k]\mathbf{G}^{-1}\mathbf{H}[n,k]\right\}=\left(1-\frac{P}{M}\right)\mathbf{I}.
\end{align}
In other words, the solution of (\ref{D2}) can be approximated by the solution of the following differential equation
\begin{align}\label{D4}
\Delta{\mathbf{U}}[n+1,k]\mathbf{G}^{\frac{1}{2}}=\left(1-\frac{P}{M}\right)\Delta{\mathbf{U}}[n,k]\mathbf{G}^{\frac{1}{2}}+\mathbf{\Phi}[n,k]\mathbf{G}^{\frac{1}{2}}.
\end{align}
Direct calculation of (\ref{D4}) yields that
\begin{align}\label{D5}
\Delta{\mathbf{U}}[n,k]\mathbf{G}^{\frac{1}{2}}=&\left(1-\frac{P}{M}\right)^n\Delta{\mathbf{U}}[0,k]\mathbf{G}^{\frac{1}{2}}+\nonumber\\
&\sum_{i=0}^{n-1}\left(1-\frac{P}{M}\right)^{n-1-i}\mathbf{\Phi}[i,k]\mathbf{G}^{\frac{1}{2}},
\end{align}
where the first term is a natural component and the second term is a forced component. Since the expansion order for initialization is assumed large enough so that $\Delta{\mathbf{U}}[0,k]=0$, (\ref{D5}) can be reduced to
\begin{align}\label{D6}
\Delta{\mathbf{U}}[n,k]\mathbf{G}^{\frac{1}{2}}=\sum_{i=0}^{n-1}\left(1-\frac{P}{M}\right)^{n-1-i}\mathbf{\Phi}[i,k]\mathbf{G}^{\frac{1}{2}},
\end{align}
or equivalently in a scalar form as
\begin{align}\label{D7}
&~~~~\sqrt{g_p}\Delta{u}_{p,m}[n,k]\nonumber\\
&=\sum_{i=0}^{n-1}\left(1-\frac{P}{M}\right)^{n-1-i}\frac{h_{p,m}^*[i+1,k]-h_{p,m}^*[i,k]}{M\sqrt{g_p}}.
\end{align}
Therefore, the MSE can be obtained as
\begin{align}\label{D8}
&\mathrm{E}\{|\sqrt{g_p}\Delta{u}_{p,m}[n,k]|^2\}=\frac{1}{M^2}\sum_{i_1=0}^{n-1}\sum_{i_2=0}^{n-1}\left(1-\frac{P}{M}\right)^{2(n-1)-(i_1+i_2)}\cdot\nonumber\\
&\left\{2J_0\left[2\pi f_d(i_1-i_2)T\right] - J_0\left[2\pi f_d(i_1-i_2+1)T\right] - \right.\nonumber\\
&\left.J_0\left[2\pi f_d(i_1-i_2-1)T\right]\right\},
\end{align}
where $J_0(\cdot)$ is the zeroth order Bessel function of the first kind. When the Doppler frequency, $f_d$, is small, we have
\begin{align}\label{D9}
J_0(2\pi f_d nT)\approx 1- \pi^2 f_d^2 n^2 T^2.
\end{align}
By substituting (\ref{D9}) into (\ref{D8}), we can obtain
\begin{align}\label{D10}
\mathrm{E}\{|\sqrt{g_p}\Delta{u}_{p,m}[n,k]|^2\}=\frac{2\pi^2f_d^2T^2}{P^2}\left[1-\left(1-\frac{P}{M}\right)^n\right]^2.
\end{align}
As a result, we have
\begin{align}
\mathrm{E}\{\|\Delta{\mathbf{U}}[n,k]\mathbf{G}^{\frac{1}{2}}\|_{\mathrm{F}}^2\}&=\sum_{m=1}^M\sum_{p=1}^P\mathrm{E}\{|\sqrt{g_p}\Delta{u}_{p,m}[n,k]|^2\}\nonumber\\
&=\frac{2\pi^2f_d^2T^2M}{P}\left[1-\left(1-\frac{P}{M}\right)^n\right]^2.
\end{align}
which is exactly (\ref{3-8_3}).

\bibliographystyle{IEEEtran}
\bibliography{IEEEabrv,massivebib}

\end{document}